\documentclass[]{article}

\pdfoutput=1
\usepackage{authblk}
\usepackage{fullpage}
\usepackage{amsmath}
\usepackage{booktabs}

\usepackage{caption}
\usepackage{graphicx}

\usepackage[hidelinks]{hyperref}

\usepackage[round, semicolon]{natbib}

\title{Extracting High-Resolution, Multi-Mode Surface Wave Dispersion Data from Distributed Acoustic Sensing Measurements using the Multichannel Analysis of Surface Waves}
\author[1]{Joseph P. Vantassel}
\author[2]{Brady R. Cox}
\author[3]{Peter G. Hubbard}
\author[1]{Michael Yust}
\affil[1]{The University of Texas at Austin}
\affil[2]{Utah State University}
\affil[3]{University of California Berkeley}

\begin{document}

\maketitle

\begin{abstract}
Distributed acoustic sensing (DAS) is a rapidly expanding tool to sense wave propagation and system deformations in many engineering applications. In terms of site characterization, DAS presents the ability to make static and dynamic strain measurements at scales (e.g., kilometers) and spatial resolutions (e.g., meters) that were previously unattainable with traditional measurement technologies. In this study, we rigorously assess the potential for extracting high-resolution, multi-mode surface wave dispersion data from DAS measurements using active-source multichannel analysis of surface waves (MASW). We have utilized both highly-controlled, broadband vibroseis shaker trucks and more-variable, narrow-band sledgehammer sources to excite the near surface, and compare the DAS-derived dispersion data obtained from both source types directly with concurrently acquired traditional geophone-derived dispersion data. We find that the differences between the two sensing approaches (i.e., DAS and geophones) are minimal and well within the dispersion uncertainty bounds associated with each individual measurement type when the following conditions are met for DAS: (a) a tight-buffered or strain-sensing fiber-optic cable is used, (b) the cable is buried in a shallow trench to enhance coupling, and (c) short gauge lengths and small channel separations are used. We also show that frequency-dependent normalization of the dispersion image following MASW processing removes the effects of scaling, integration, and differentiation on the measured waveforms, thereby allowing nearly identical dispersion data to be extracted from geophone waveforms (proportional to velocity) and DAS waveforms (proportional to strain) without requiring them to first be converted into equivalent units. We provide evidence that the short wavelength (high frequency) DAS dispersion measurements are limited by both the gauge length and the more commonly considered channel separation. We further show that it is possible to extract essentially equivalent surface wave dispersion data from seismic measurements made using a traditional geophone array or two different DAS cables. Finally, we show that shear wave velocity profiles recovered from the DAS data using an uncertainty-consistent, multi-mode inversion agree favorably with cone penetration tests performed at the site. This study demonstrates that DAS, when appropriate considerations are made, can be used in-lieu of traditional sensors (i.e., geophones) for making high-resolution, multi-mode measurements of surface wave dispersion data using the MASW technique.
\end{abstract}

\pagebreak

\section{Introduction}

Distributed acoustic sensing (DAS) is an emerging technology with wide applications in infrastructure health monitoring and site characterization \citep{hubbard_dynamic_2021, hubbard_road_2022, lindsey_cityscale_2020, spikes_comparison_2019, wang_ground_2018}. DAS permits the acquisition of static and dynamic signals at length scales (e.g., kilometers) and spatial resolutions (e.g., meters) previously unattainable with traditional sensing technologies \citep{soga_distributed_2018}. DAS, and for that matter the larger area of distributed fiber optic sensing (DFOS), requires three main system components: (1) the fiber-optic cable, (2) the interrogator unit (IU), and (3) dedicated storage, computation, and visualization resources. The fiber-optic cable is the sensing instrument whose elongation or compression (i.e., strain) is measured by the DAS system. Fiber-optic cables are specially designed to propagate the light emitted by the interrogator unit with minimal loss, allowing light to travel (and therefore strain measurements to be made) over large distances (i.e., tens of kilometres) \citep{lindsey_fiber-optic_2021}. Note that the fiber must be selected and installed carefully to ensure acceptable results, however, as even specially designed fiber-optic cables for strain-sensing applications are relatively inexpensive (between \$3 and \$7 per meter) they are typically not retrieved for re-use after testing concludes. The second component, the interrogator unit, is connected to one end of the fiber-optic cable for the purpose of sending pulses of light down the length of the fiber and measuring the returned reflections. The reflections that occur primarily through the back-scattering of light inside of the cable are, through the use of precise timing and fast sampling rates (i.e., 100s of kHz), interpreted by the IU into measurements of phase change along the fiber \citep{karrenbach_fiber-optic_2019}. The phase change measurements made by the IU can then be transformed into one-dimensional measurements of strain at various physical distances along the fiber. The IU is by far the most expensive piece of equipment required for DAS, with high-quality units ranging in price between approximately \$150k and \$500k. Dedicated storage, computational, and visualization resources are typically a high-end computer, with large amounts of dedicated storage (at least multiple terabytes), and a real-time data acquisition interface to facilitate high-quality DAS measurements. Note that for many systems the IU and dedicated storage, computation, and visualization resources are combined into a single unit capable of both acquiring and storing the data and providing easily understood feedback to the experimenters.

The multichannel analysis of surface waves (MASW) is an active-source surface-wave testing technique for estimating a site's surface wave dispersion from recordings of dynamic signals with strong surface wave content \citep{park_multichannel_1999}. The MASW technique most commonly involves the utilization of a linear array of receivers, typically velocity transducers (i.e., geophones), and a surface wave source located collinear with the array and operated by the experimenters. Geophones are most commonly oriented vertically, but may also be oriented horizontally in the in-line or cross-line directions (i.e., sensing particle motion collinear with or perpendicular to the array, respectively). Just as with the geophones, the seismic source may be oriented vertically, horizontally in-line, or horizontally cross-line, depending on the goals of testing. The most common configurations involve the utilization of a vertical source with vertical receivers to measure Rayleigh-type surface wave dispersion and a horizontal cross-line source with horizontal cross-line receivers to measure Love-type surface wave dispersion \citep{vantassel_swprocess_2022}. In this study, we will use a less common experimental setup that utilizes both vertical and horizontal in-line sources acquired on horizontal in-line receivers to measure Rayleigh-type surface wave dispersion. 

While the vast majority of prior studies involving DAS and surface wave testing have focused on the use of DAS for passive-wavefield/ambient-noise measurements (e.g., Luo et al., \citeyear{luo_horizontally_2020}; Shragge et al., \citeyear{shragge_low-frequency_2021}; Zeng et al., \citeyear{zeng_properties_2017}) few have investigated the use of DAS for active-source surface wave testing. One of the earliest published studies of using DAS for active-source surface wave testing is Galan-Comas (\citeyear{galan-comas_multichannel_2015}). In their study, they extract fundamental Rayleigh mode surface wave dispersion data from active-source DAS and geophone wavefield recordings made using a sledgehammer source. The DAS acquisition system consisted of a coherent OTDR interrogator (CR3 Prototype System by Optiphase\textsuperscript{\textregistered}) with a buried standard telecommunications cable (SST-Ribbon\textsuperscript{TM} Single-Tube, Gel-Free, Armored Cable, 24 F, Single-mode (OS2)). The DAS system utilized a 5-m gauge length and a 2-m channel separation. The adjacently deployed geophone array consisted of 4.5 Hz geophones deployed at a 1-m spacing. Both systems were deployed over approximately 71 m. While the DAS- and geophone-derived dispersion agreed well (i.e., within the typical range of dispersion uncertainty 5 – 10 \%), the DAS-derived dispersion was bandlimited, only extending between 6 and 20 Hz (wavelengths of approximately 40 and 8 m, respectively), compared to the geophone data, which extended between 6 and 56 Hz (wavelengths of approximately 40 and 2.5 m, respectively). Later, Costley et al. (\citeyear{costley_spectral_2018}) used the spectral analysis of surface waves (SASW) technique to compare effective mode Rayleigh surface wave dispersion from DAS and geophone data generated using a sledgehammer source. The DAS acquisition system consisted of an interrogator built by the Optical Techniques Branch of the Naval Research Lab with a buried, unarmoured, loose tube, gel-filled fiber-optic cable manufactured by OFS. The DAS system utilized a 10-m gauge length and a 10-m channel separation. The geophone array for SASW consisted of 4.5 Hz geophones spaced between 10 and 40 m. The quality of the resulting effective mode dispersion comparisons varied considerably, from good (with 5 – 10 \%) to poor (up to 50 \%), depending upon the segment of the approximately 40 m-long array that was processed. Song et al. (\citeyear{song_imaging_2018}) used MASW to extract fundamental mode Rayleigh surface wave dispersion using active-source cross-correlated seismic records generated by a vibroseis shaker truck source. While neither the DAS acquisition system nor the fiber-optic cable are described in the study, it is noted that the DAS system used a 10-m gauge length and 1-m channel separation and that the fiber-optic cable was buried. The linear segment of the much larger two-dimensional array was approximately 135 m in length. Fundamental mode Rayleigh dispersion data for both the DAS and geophone measurements were extracted over a relatively narrow band (5 – 20 Hz, wavelengths of 120 and 15 m, respectively), although based on the results presented the geophone-derived dispersion was of good quality at 20 Hz and could likely have been extracted to high-frequencies (shorter wavelengths). The DAS- and geophone-derived dispersion data were found to be within approximately 25\% of each other (i.e., well outside of the typical dispersion uncertainty range), with the geophone data tending to predict higher surface wave phase velocities than the DAS. The less than ideal comparison was primarily attributed to a large geophone receiver spacing (40 m), resulting in significant spatial aliasing during MASW processing of the geophone measurements. The limited and mixed results observed in these prior studies requires a more detailed and comprehensive comparison of the ability of DAS to be used for active-source surface wave testing.

In this paper, we compare the abilities of DAS and traditional geophone sensors to acquire active-source dynamic signals for the purpose of extracting high-resolution, multimode surface wave dispersion data using the MASW technique. This study utilizes a 94-m section of two different 200-m long fiber-optic cables and an adjacently deployed 94-m long geophone array (48 receivers at a 2-m spacing). The geophone and DAS arrays were simultaneously used to record dynamic signals rich in surface wave energy generated by off-end vibroseis shaker truck and sledgehammer impulse sources. The DAS-derived and geophone-derived waveforms were then used in an MASW workflow to extract multi-mode dispersion data. The dispersion extracted from the geophone and DAS measurement systems show excellent agreement with one another, and shear wave velocity (Vs) profiles derived from the experimental dispersion data using a multi-mode uncertainty-consistent inversion procedure are compared to cone penetration tests performed along the array. This study demonstrates that when appropriate considerations are made, DAS can be used as a replacement for traditional geophone deployments to extract high-resolution, multi-mode surface wave dispersion data and recover meaningful subsurface Vs profiles.

\section{Experimental Setup}

\begin{figure}[t]
	\includegraphics[width=\textwidth]{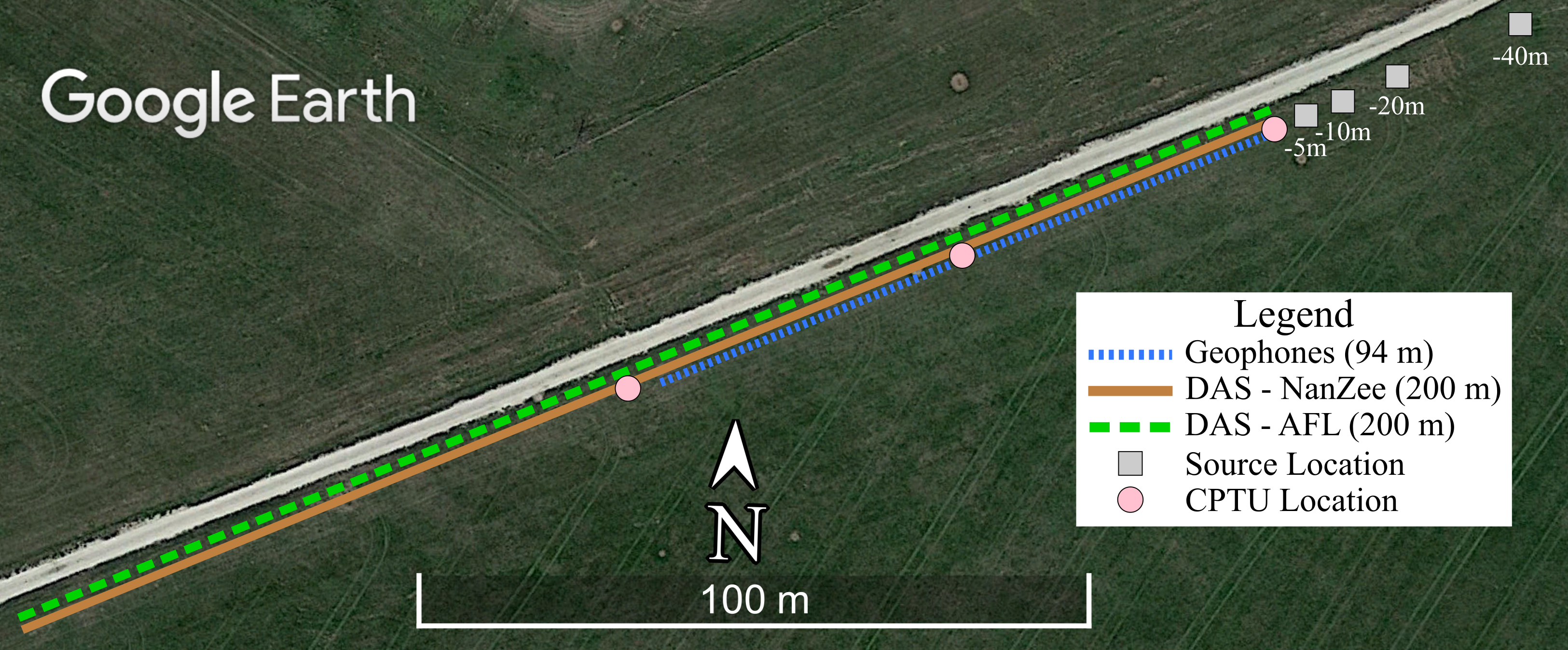}
	\caption{Plan view of the experimental setup at the Hornsby Bend test site in Austin, Texas, USA, where a 94-m long geophone array (48 horizontal receivers oriented inline at a 2-m spacing) was deployed alongside 200 m of NanZee and 200 m of AFL fiber-optic cable to compare surface wave dispersion data extracted from geophone and DAS waveforms. Surface wave energy was produced from four distinct shot locations denoted as -5 m, -10 m, -20 m, and -40 m. To provide a source of ground truth for the very near-surface (depths $<$ 10 m) three cone penetration tests with pore pressure readings (CPTU) were made at approximately 0, 50, and 100 m along the array.}
	\label{fig:1}
\end{figure}

\begin{figure}[t]
	\centering
	\includegraphics[width=0.5\textwidth]{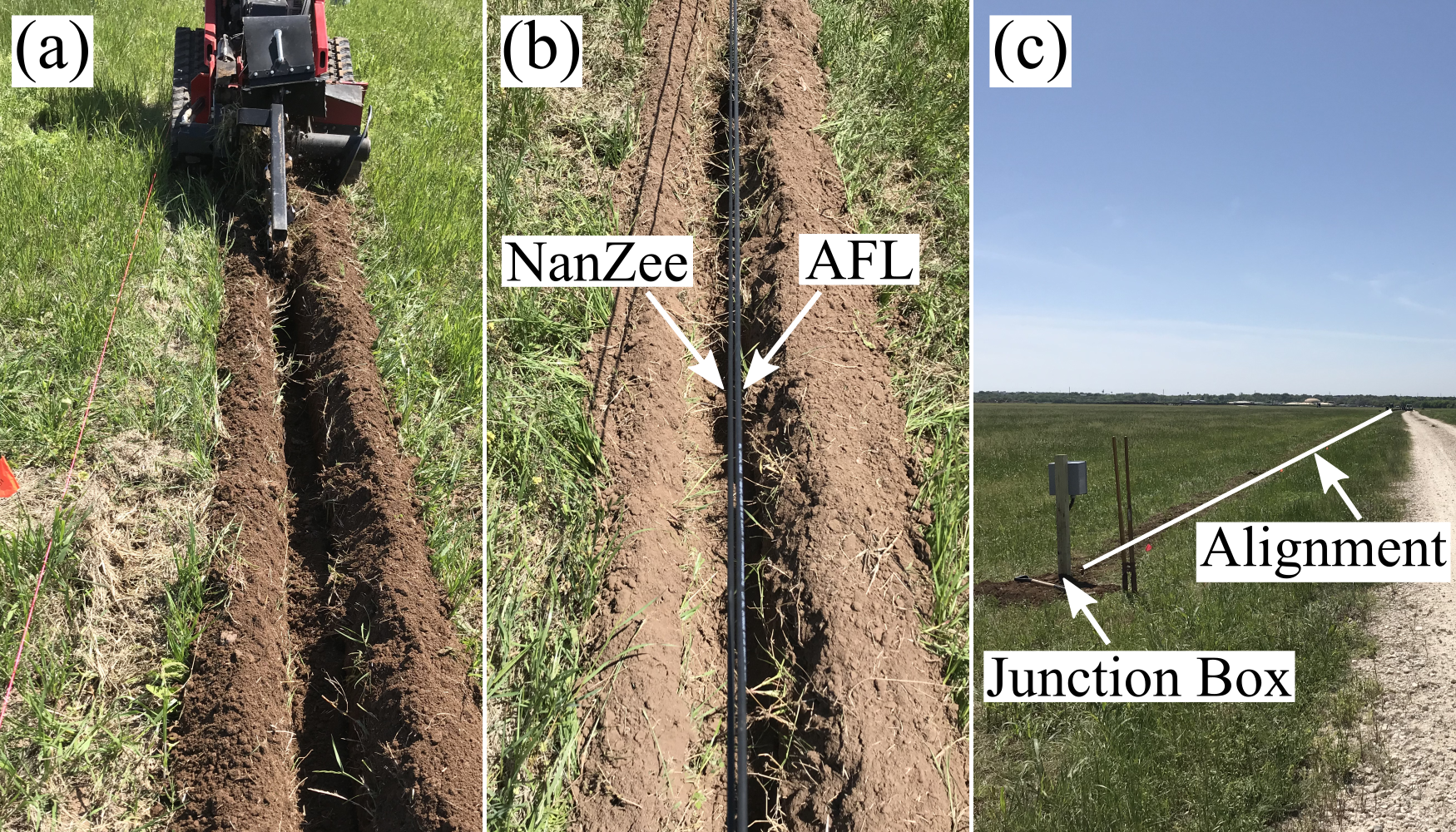}
	\caption{The installation procedure for the NanZee and AFL fiber-optic cables involved: (a) using a trenching machine to excavate a trough between approximately 10 and 15 cm deep, (b) placing the two fiber-optic cables side-by-side in the trench, and (c) back-filling and compacting the trench while ensuring the cables remained in alignment and bringing the ends of the fiber-optic cable up to junction boxes located at either end of the array.}
	\label{fig:2}
	\centering
\end{figure}

The experiment was conducted at the NHERI@UTexas \citep{stokoe_nheriutexas_2020} Hornsby Bend test site in Austin, Texas, USA. A plan view of the experimental setup is shown in Figure \ref{fig:1}. For this experiment, a 94-m geophone array (48 receivers at 2-m spacing) was deployed alongside two 200-meter long fiber-optic cables. The two fiber-optic cables were deployed in the same shallow trench at the site; one from NanZee Sensing Technology (NZS-DSS-C02) and the other from AFL (X3004955180H-RD). Importantly, both cables were constructed such that strain is transferred from the exterior of the cable to the fiber-optic core. This is achieved through a textured exterior surface and a tightly-buffered interior construction  \citep{soga_distributed_2018}. In addition to proper cable selection, the cable must be installed carefully to ensure effective transmission of strain from the ground into the cable's exterior. The cable installation process is documented in Figure \ref{fig:2}. The linear fiber-optic array was first located by surveying its position using a total station. Second, a trenching machine was used to excavate a shallow trench between 10 and 15 cm deep in which to place the cables (see Figure \ref{fig:2}a). Both cables were then installed in the trench adjacent to one another (see Figure \ref{fig:2}b). Finally, the ends of the fiber-optic cable were brought up to junction boxes located at either end of the array, the cables were slightly tensioned to assure minimal slack, and the trench was carefully backfilled and compacted to ensure good coupling of the cable with the surrounding soil (see Figure \ref{fig:2}c). The two cables were joined together at the far end of the array by splicing the NanZee and AFL cables to one another. This allowed for simultaneous recording on both cables. On the near-side of the array, the NanZee fiber was connected to the IU, which for this experiment was an OptaSense ODH4, and the AFL fiber was appropriately terminated to reduce end-reflections. The ODH4 IU was configured by a team from OptaSense to ensure high-quality data acquisition. As the OptaSense ODH4 allows for a variable gauge length, the shortest possible gauge length of 2.04 m was selected for these experiments. For those readers who may not be familiar, the gauge length represents the length of fiber that the elongation (or strain) is measured over. Effectively, the 2.04-m gauge length used in this study means that at each sampling location (or channel separation) the resulting vibrations will be an average over that 2.04-m gauge length. The effect of gauge length on surface wave dispersion will be discussed later in this work. A 1.02-m channel separation, the shortest allowed by the ODH4, was selected. The short channel separations allowed waveform measurements every 1.02 m along the cable, however, these waveforms represented an average response over the 2.04-m gauge length surrounding each channel location. Note that in the remainder of the work the channel spacing and gauge length will be rounded to the nearest meter for simplicity. The sampling frequency (also known as the ping rate) of the interrogator unit was set at 100 kHz. Oversampling, that is sampling faster than required by the Nyquist sampling theorem, serves to improve the measurement's resolution (i.e., effectively increase the system's dynamic range) and signal-to-noise (SNR). Following acquisition, and prior to dispersion processing, the raw measurements were down sampled to 1 kHz and high-pass filtered above 3 Hz to remove artifacts at low frequencies that consist of laser drift and static strains. The channels that corresponded to the first and last sensing segments of the two DAS cables were identified and mapped to physical locations by using a modified version of tap testing, wherein an off-end source impact was used excite the cable and later used to identify the closest channel with significant coherent wave energy. This process was repeated on both sides of the array to identify the first and last sensing segment of each cable.

Immediately adjacent to the fiber-optic cables, two geophone arrays (one vertical and one horizontal in-line), each consisting of 48 receivers, were deployed. To allow for direct comparison with the DAS arrays, which are sensing in the horizontal in-line direction, only the results from the horizontal in-line geophones will be discussed in this study. The vertical and horizontal geophones were from Geospace Technologies (GS-11D) and had a resonant frequency of 4.5 Hz. The geophones were mounted in PC21 land cases and coupled to the ground surface with 7.6-cm aluminum spikes. The vertical and horizontal geophones were deployed at a constant 2-m spacing, resulting in a total array length of 94 m. Note that a longer geophone array equal in length to the DAS cables (i.e., 200 m) could not be deployed due to equipment constraints, which would have required 96 geophones for each geophone array (vertical and horizontal in-line) to maintain a 2-m spacing. This highlights one of the key advantages of DAS in that a single IU with a single fiber-optic cable can acquire signals over distances (up to several kilometers) and at a spatial resolution (meter-scale) that would be infeasible with traditional geophones. Signals from both the vertical and horizontal geophone arrays were recorded simultaneously using four interconnected 24-channel Geometric Geode seismographs. All signals were acquired using a sampling rate of 1 kHz. To provide a source of ground truth for the very near-surface (depths $<$ 10 m) at the Hornsby Bend test site, three cone penetration tests with pore pressure measurements (CPTUs) were made at approximately 0, 50, and 100 m along the array. The location of these tests and their proximity to the array's alignment are shown in Figure \ref{fig:1}.

\section{Data Acquisition}

\begin{figure}[t]
	\includegraphics[width=\textwidth]{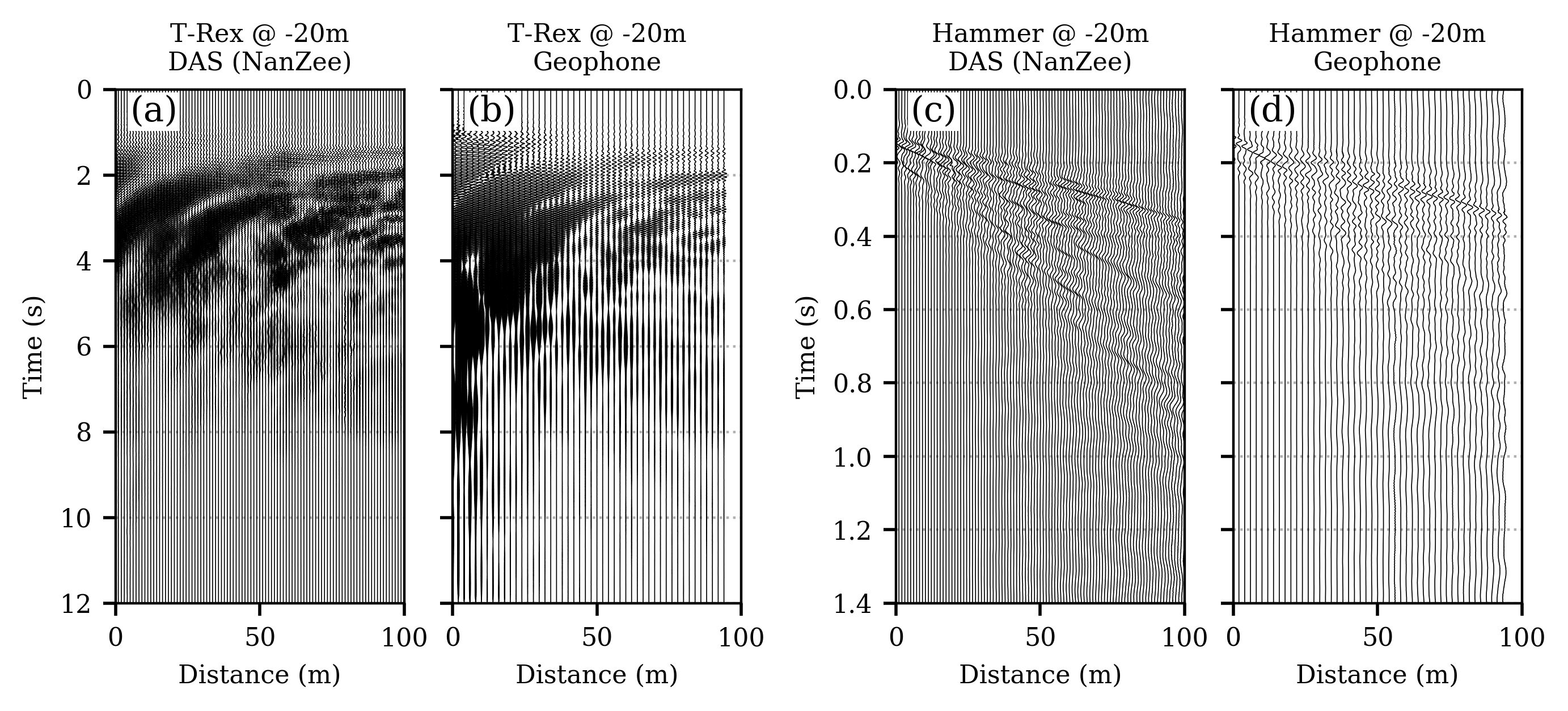}
	\caption{Example waveforms from the Hornsby Bend test site. The waveforms are the result of T-Rex shaking horizontally inline at -20 m, as recorded on the (a) NanZee cable and (b) the horizontal geophone array, and vertical sledgehammer impacts at -20 m as recorded on the (c) NanZee cable and (d) the horizontal geophone array. The waveforms shown have been stacked in the time-domain.}
	\label{fig:3}
\end{figure}

The geophone array and DAS fiber-optic cables were used to simultaneously record actively-generated surface waves from different sources, including highly-controlled vibroseis shaker trucks and more-variable impact sources. The vibroseis sources include the specialized three-dimensional shaker T-Rex and the highly-mobile one-dimensional shaker Thumper from the NHERI@UTexas experimental facility \citep{stokoe_nheriutexas_2020}. T-Rex was used to shake in all three directions (i.e., vertically, horizontally in-line, and horizontally cross-line), however, only the vertical and horizontal in-line shakes will be discussed in this work. T-Rex was used to produce a 12 second chirp with frequencies swept linearly from 3 to 80 Hz. T-Rex has a maximum force output of approximately 270 kN in the vertical and 130 kN in the horizontal directions. The other vibroseis source, Thumper, was used to produce a 12 second chirp in the vertical direction with frequencies swept up linearly from 5 to 200 Hz. Thumper has a maximum force output in the vertical direction of approximately 27 kN. The impact source used for this study was an instrumented 5.4 kg sledgehammer from PCB Piezotronics. The frequency content produced by the sledgehammer is highly variable and is dependent on the operator, the strike plate used, and the material being tested (e.g., stiffer materials tend to illicit higher frequencies). The peak force output of the sledgehammer from the tests performed at Hornsby Bend was approximately 20 kN. All four of the sources (i.e., T-Rex shaking vertically, T-Rex shaking horizontally in-line, Thumper shaking vertically, and the sledgehammer striking vertically) were deployed at various shot locations around the site. However, for this study we will focus only on the source positions located at 5, 10, 20, and 40 m away from the start of the geophone and DAS arrays. The source positions relative to the array are shown in Figure \ref{fig:1}. For the vibroseis sources, three sweeps were performed at each source location, whereas for the sledgehammer source five impacts were performed. An example of waveforms stacked in the time domain, one from T-Rex shaking horizontally inline at -20 m and one from a vertical sledgehammer impact at -20 m, as recorded on one of the DAS cables (i.e., NanZee) and the horizontal geophone array, are shown in Figure \ref{fig:3}. The raw seismic wavefield recordings made with the DAS and geophone arrays have been made publically available on the DesignSafe-CI \citep{vantassel_active-source_2022}.

\section{Effect of Measurement Domain}

Before comparing the surface wave dispersion data extracted from the geophone and DAS arrays, we must first briefly consider the effect of processing waveforms in different units; namely, waveforms in terms of particle velocity and particle displacement. This is critical for the present study, as the geophone measurements are proportional to velocity, whereas the DAS measurements made in the course of this study are proportional to strain (note that some DAS systems acquire data proportional to strain-rate). Note that the optimal procedure for converting the geophone and DAS waveforms to consistent physical units (i.e., particle velocity or particle displacement) is non-trivial and an area of on-going research.

\begin{figure}[t]
	\includegraphics[width=\textwidth]{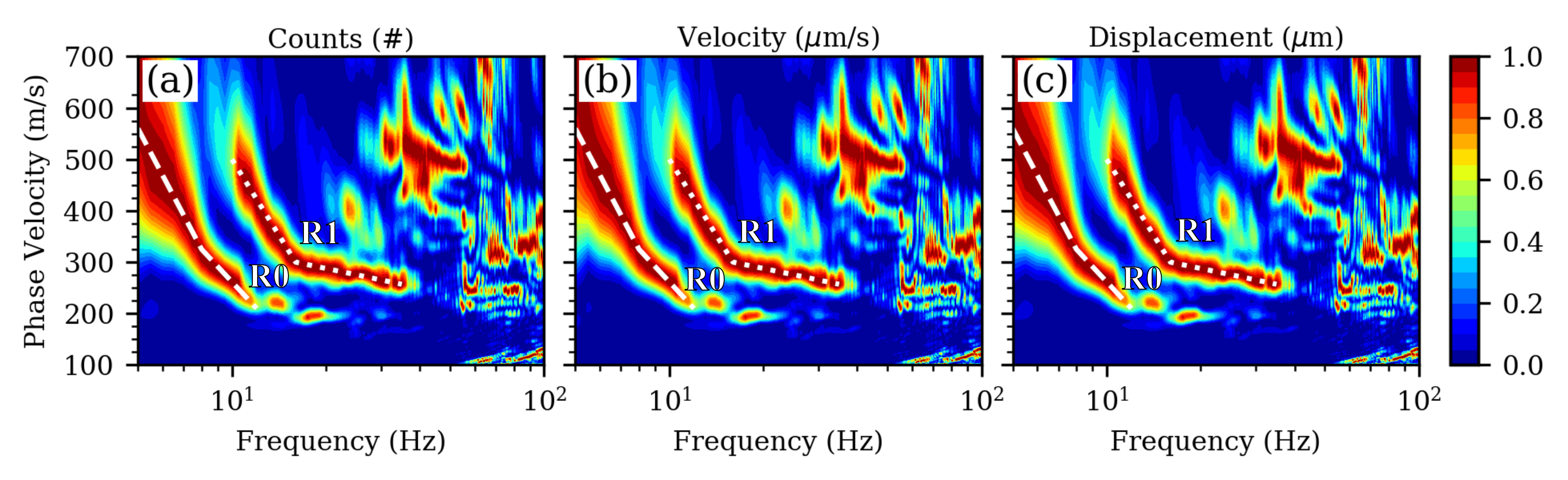}
	\caption{Comparison of surface wave dispersion images derived from three vertical T-Rex chirps stacked in the time domain at the source location of -20 m using waveforms recorded by the geophone array and processed in terms of: (a) raw counts, (b) particle velocity in units of micrometers per second, and (c) particle displacement in units of micrometers. Note that the use of frequency-dependent normalization eliminates the requirement of converting the geophone-derived and, by extension, the DAS-derived waveform data to consistent physical units prior to performing MASW processing. To allow easy comparisons between the dispersion images, the approximate trend of the apparent fundamental and first-higher Rayleigh wave modes (R0 and R1) from panel (a) are shown in panels (b) and (c) using a dashed and dotted line, respectively. The wavefield transformation shown is the frequency-domain beamformer (FDBF) with cylindrical-steering vector and square-root weighting. Warm and cool colors represent high and low relative surface wave power, respectively.}
	\label{fig:4}
\end{figure}

Therefore, it is desirable to first assess the effect of performing dispersion processing using several geophone-derived wavefields in different units. Figure \ref{fig:4} shows dispersion images extracted from a seismic wavefield produced by time-domain stacking of three vibroseis sweeps generated by T-Rex shaking vertically at the -20 m source position when the waveforms were processed in terms of: (a) raw units of counts, (b) converted to particle velocity, and (c) numerically integrated to particle displacement. The dispersion images were produced using the frequency-domain beamformer (FDBF) with cylindrical-steering vector and square-root weighting \citep{zywicki_mitigation_2005}, as implemented in the open-source software \emph{swprocess} \citep{vantassel_jpvantasselswprocess_2021}. Following frequency domain beamforming, each dispersion image was normalized by its maximum power at each frequency and contoured to produce the dispersion images shown in Figure \ref{fig:4}. To show the similarity between the three dispersion images, the apparent fundamental and first-higher Rayleigh modes (R0 and R1) from panel (a), delineated using dashed and dotted lines, respectively, are also plotted in panels (b) and (c). The recovered modes in all three domains show excellent agreement and demonstrate that frequency-dependent amplitude normalization of the dispersion image is an effective technique for eliminating the effect of processing surface wave dispersion data using different units. Note that the ability of frequency-dependent normalization to remove the effects of scaling and integration is consistent with the underlying mathematics. In short, when scaling a time series by a constant (e.g., transforming from counts to voltage), the amplitude at each frequency also scales by a constant. Similarly, when performing numerical integration (e.g., transforming from particle velocity to particle displacement), which is equivalent to dividing the Fourier amplitudes at each frequency by the factor $j\omega$, the Fourier amplitudes at each frequency are also scaled. Since, frequency-dependent normalization remove these scaling effects, we can confidently extend these same principles to the DAS-derived waveforms, thereby allowing us to compare geophone-derived and DAS-derived dispersion data by processing each measurement in their most raw form without the need to first convert into consistent engineering units.

\section{Surface Wave Dispersion from the Raw Seismic Wavefields}

\begin{figure}[t]
	\includegraphics[width=\textwidth]{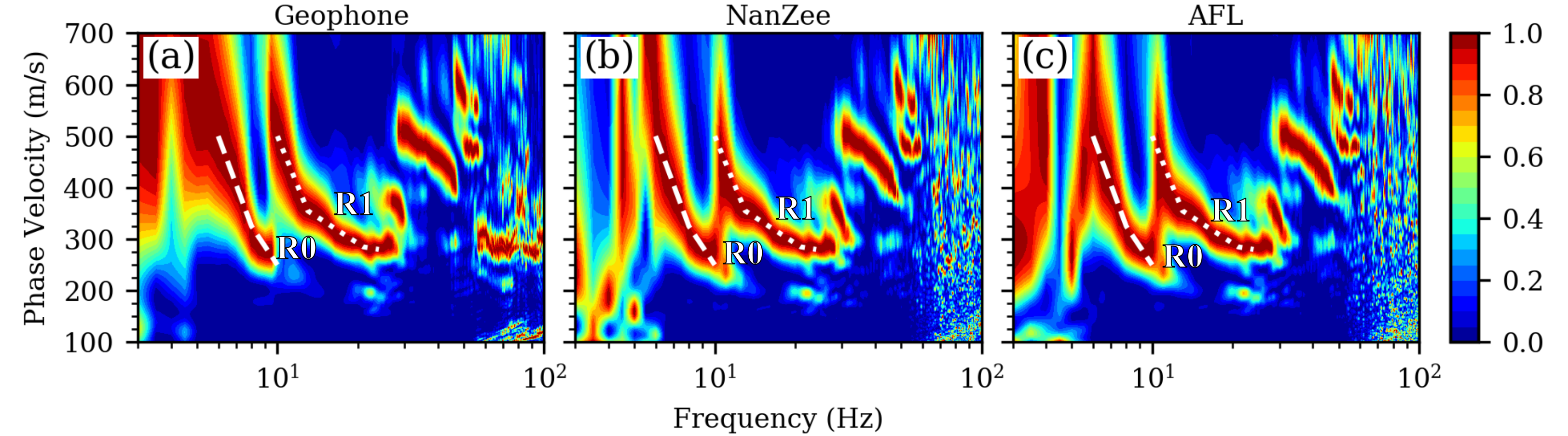}
	\caption{Comparison of surface wave dispersion images from three vertical T-Rex chirps stacked in the time domain at the source location of -40 m as derived from the seismic wavefield measured by the (a) geophone array, (b) NanZee cable, and (c) AFL cable. The wavefield transformation shown is the FDBF with cylindrical-steering vector and square-root weighting. To easily compare the dispersion images, an approximate trend of the apparent fundamental and first-higher Rayleigh wave modes (R0 and R1) from panel (a) are shown in all panels using a dashed and dotted line, respectively. Warm and cool colors represent high and low relative surface wave power, respectively.}
	\label{fig:5}
\end{figure}

With the effect of different measurement domains mitigated through frequency-dependent normalization, we now present a comparison of surface wave dispersion images recovered from the raw seismic wavefields. Figure \ref{fig:5}a, \ref{fig:5}b, and \ref{fig:5}c compares the surface wave dispersion images for T-Rex shaking vertically at -40 m as derived from the raw wavefield recorded on the geophone array, NanZee cable, and AFL cable, respectively. The wavefield transformation used is again the FDBF with cylindrical-steering vector and square-root weighting. To ensure a fair comparison with the geophone array, only the first 94 m of the DAS arrays were processed. To easily compare the dispersion images, an approximate trend of the apparent fundamental and first-higher Rayleigh wave modes (R0 and R1) from panel (a) are shown in all panels using a dashed and dotted line, respectively. A visual comparison of the three dispersion images reveal excellent agreement between the three acquisition systems with R0 and R1 being particularly clear. The dispersion images also contain other high-relative-power trends that are less clear than R0 and R1 but potentially indicative of other higher modes. Similar to the observations made in regard to Figure \ref{fig:4}, we see at low and high frequencies, $<$6 Hz and $>$60 Hz for this source and offset combination that the dispersion images consist of a considerable amount of incoherent noise. However, despite the noise at high and low frequencies, the authors wish to emphasize the excellent consistency observed between the geophone and DAS arrays.

\section{Effect of Trace Separation and Gauge Length}

As mentioned previously, there are two key parameters of the DAS experimental setup that impact spatial resolution: the IU's channel separation and gauge length. The purpose of this section is to summarize the effects and relative importance of these two acquisition parameters for near-surface site characterization. We begin with the more intuitive of the two parameters, the channel separation. The channel separation (i.e., the distance between DAS readings along the fiber-optic cable) is analogous to the receiver spacing in traditional array measurements. A shorter channel separation generally allows for the acquisition of shorter wavelengths, however, a shorter channel separation (over the same array length) will produces greater quantities of data, requiring additional storage and processing. Therefore, the channel separation should not always be set as small as possible, but rather to a value that provides adequate spatial sampling without the acquisition of excess data. The minimum channel separation for adequate spatial sampling requires at least two channels per wavelength for the shortest wavelength of interest (the Nyquist sampling theorem applied in space). We do note that for surface wave methods, previous work \citep{foti_guidelines_2018, vantassel_swprocess_2022} has shown  that strict adherence to the two samples per minimum wavelength criterion is not strictly necessary if a clear dispersion trend can be identified and appropriate precautions are taken to avoid spatial aliasing. Through the present study, the authors have seen no indication that this possible relaxation of the Nyquist sampling theorem in space does not also apply to measurements made with DAS, although in practice we recommend ensuring at least two samples per minimum wavelength whenever possible to ensure the highest quality data acquisition.

\begin{figure}[t]
 	\includegraphics[width=\textwidth]{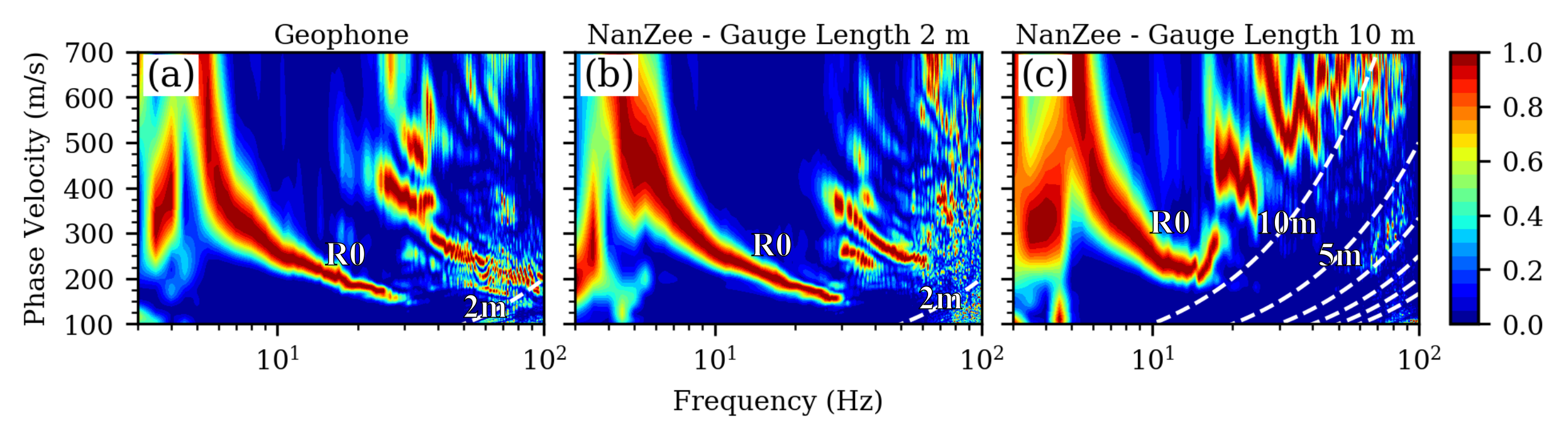}
 	\caption{Comparison of surface wave dispersion images from three in-line T-Rex chirps stacked in the time domain at the source location of -5 m, as derived from the seismic wavefield measured by the: (a) geophone array, (b) NanZee cable with a 2-m gauge length and 1-m channel separation, and (c) NanZee cable with a 10-m gauge length and 1-m channel separation. The wavefield transformation shown is the FDBF with cylindrical-steering vector and square-root weighting. The fundamental Rayleigh mode (R0) is denoted in all panels. Dashed light-colored lines are drawn in all panels to indicate wavelengths of importance. In panels (a) and (b) a dashed line denotes a wavelength of 2 m. In panel (c) the dashed line indicates wavelengths equal to the gauge length (10 m for this panel) divided by n, where n is a positive integer. However, only values up to n = 6 are shown for simplicity. Warm and cool colors represent high and low relative surface wave power, respectively.}
 	\label{fig:6}
 \end{figure}

The second and more complex of the two acquisition parameters is the gauge length. The gauge length represents the length of cable over which strain is measured, and effectively averaged, by the IU. Importantly, there is no direct analogy between gauge length and traditional geophone-based array measurements, as geophones make a discrete point measurement (i.e., do not average in space). Intuitively, the larger the gauge length the more spatial averaging that will occur. Therefore, we should expect that for DAS, wavelengths that are short relative to the gauge length will be averaged over the entire gauge length, resulting in the loss of information, whereas, wavelengths that are long relative to the gauge length will be well recovered. To illustrate how the averaging effect of gauge length impacts measurements of surface wave dispersion, Figure \ref{fig:6}a, \ref{fig:6}b, and \ref{fig:6}c present dispersion images from in-line T-Rex shaking 5 m away from the sensing arrays, as measured by the geophone array with a 2-m receiver spacing, the NanZee cable with a 2-m gauge length and 1-m channel separation, and the NanZee cable with a 10-m gauge length and a 1-m channel separation, respectively. The wavefield transformation shown is the FDBF with cylindrical-steering vector and square-root weighting. A visual comparison of the R0 trend from the geophone array (Figure \ref{fig:6}a) and NanZee cable with a 2-m gauge length (Figure \ref{fig:6}b) reveals good agreement. In fact, it could be argued that the NanZee cable with a 2-m gauge length provides a clearer R0 trend at low frequencies and more coherent higher mode trends at high frequencies, which could be related to the 1-m channel separation. Importantly, we do not observe any significant wavelength-limiting, gauge-length-related effects in the DAS dispersion image relative to the image obtained from the 2-m geophone array. In contrast, if we compare the first two dispersion images with the dispersion image extracted from the NanZee cable with a 10-m gauge length (Figure \ref{fig:6}c), we see clear evidence of a wavelength-limiting, gauge-length-related effect, such that we cannot observe any clear dispersion trends at wavelengths less than about the gauge length (i.e., 10 m). For reference, several different curved lines of constant wavelength are indicated in the figures by dashed white lines. The reader will note that the coherent dispersion trends in Figure \ref{fig:6}c disappear for wavelengths that are shorter than approximately 10 m, which is equivalent to the gauge length.

\begin{figure}[t]
	\includegraphics[width=\textwidth]{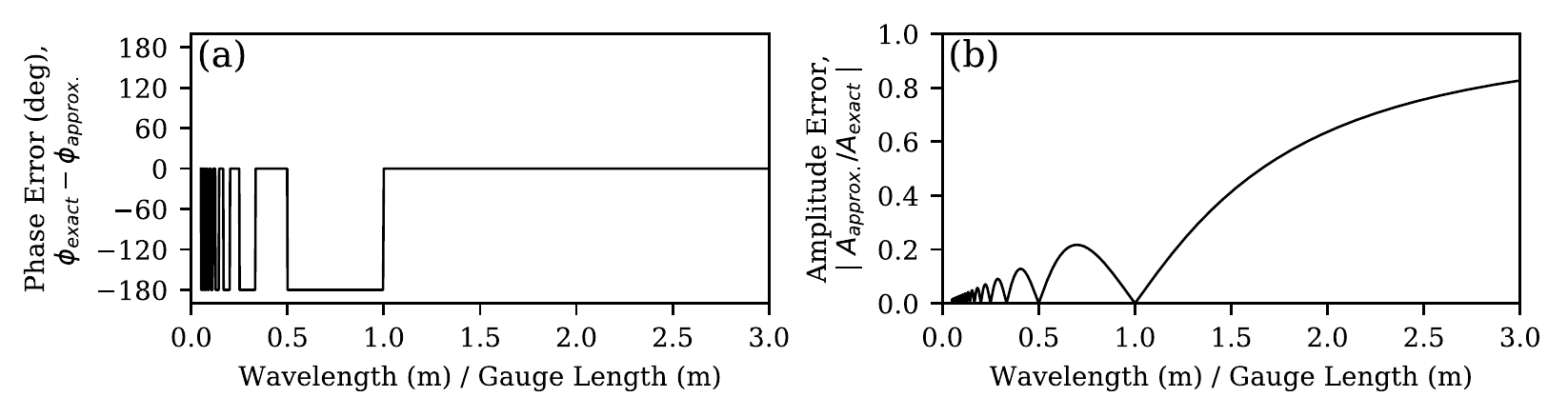}
	\caption{Effect of wavelength ($\lambda$) and gauge length ($gl$) on the: (a) phase, and (b) amplitude error in the measured wavefield. The error in phase is taken as the difference between the true phase ($\phi_{exact}$) and the approximate phase ($\phi_{approx.}$) as would be measured by DAS. The error in amplitude is taken as the absolute value of the ratio between the approximate amplitude ($A_{approx.}$) as would be measured by DAS and the exact amplitude ($A_{exact}$). The relationships presented here assume that the channel separation permits at least two samples per $\lambda$ to ensure proper spatial sampling.}
	\label{fig:7}
\end{figure}

We now provide a rigorous, physics-based explanation of the gauge length limitation phenomena observed in Figure \ref{fig:6}. We start by assuming a sinusoidal displacement wave traveling down the cable of the form:

\begin{equation}
	u(x,t) = A sin(\omega t - k x + \phi)
	\label{eq:1}
\end{equation}

where $x$ is the distance along the array, $t$ is time, $A$ is the wave's amplitude, $k$ is the wave's wavenumber, $\omega$ is the wave's circular frequency, and $\phi$ is the wave's initial phase angle. From the wave's displacement we can calculate the exact strain wavefield along the cable as:

\begin{equation}
	strain_{exact}(x,t) = \frac{\partial u(x,t)}{\partial x} = -A k cos(\omega t - k x + \phi)
	\label{eq:2}
\end{equation}

where all terms have been defined previously. However, as DAS does not make point measurements, but rather takes an average over the gauge length, we must integrate the true strain wavefield over a gauge length at each discrete channel location. Or, we can also equivalently take the finite difference in terms of displacement along the gauge length as:

\begin{equation}
	strain_{DAS}(x,t) = \frac{1}{gl} \int_{-gl/2}^{gl/2} \frac{\partial u(x,t)}{\partial x} dx = \frac{u(x+gl/2, t) - u(x-gl/2, t)}{gl}
	\label{eq:3}
\end{equation}

where $gl$ is the gauge length and all other terms have been defined previously. Rewriting Equation \ref{eq:3} we can show,

\begin{equation}
	strain_{DAS}(x,t) = \frac{2A}{gl} (cos(\omega t - k x + \phi) sin(\frac{-k gl}{2}))
	\label{eq:4}
\end{equation}

where all terms have been defined previously. To ascertain the relative effect of measuring the strain wavefield using DAS, we take the ratio of the strain as measured by DAS and the true strain:

\begin{equation}
	\frac{strain_{DAS}}{strain_{exact}} = \frac{-2}{k gl} sin(\frac{-k gl}{2}) = \frac{\lambda}{\pi gl} sin(\frac{\pi gl}{\lambda})) 
	\label{eq:5}
\end{equation}

where $\lambda$ is the wave's wavelength (i.e., $\lambda = \frac{2 \pi}{k}$) and all other terms have been defined previously. Note that from Equation \ref{eq:5} the effect of sampling the wavefield with DAS is only a function of the ratio between gauge length and wavelength. To better understand the implications of Equation \ref{eq:5}, Figure \ref{fig:7} presents the error in the measured wave's phase (Figure \ref{fig:7}a) and amplitude (Figure \ref{fig:7}b). From Figure \ref{fig:7}, for wavelengths less than one gauge length there is a change in sign (180 degree phase change) and a significant drop in amplitude ($<$ 20 \% of the true strain). The combination of these effects provide a physical basis for the prior experimental observations made in regard to Figure \ref{fig:6}. Given this theoretical backing, we observe that between a wavelength of gauge length/2 and gauge length/3 we do not have a change in sign/phase, so theoretically the correct phase angle (and therefore the correct dispersion) could be resolved. However, there is only a small strain amplitude (~10 \% of the true strain) in this wavelength region, which could make it difficult to accurately sense these waves if there is significant noise. To investigate the possibility of sensing waves with wavelengths shorter than the gauge length, additional dashed lines are provided in Figure \ref{fig:6}c that denote the locations of the sign/phase changes from Figure \ref{fig:7}a. As expected, the first region between wavelengths of one and one half of a gauge length (10 and 5 m respectively) presents no observable dispersion trends, however, for the second region between wavelengths of one half and one third of a gauge length (5 and 3.3 m, respectively) we do see some dispersive energy, although it is incoherent. This indicates that while it may be possible in theory to resolve some wavelengths less than the gauge length, extracting them in practice will likely prove challenging due to their low amplitudes. In summary, the gauge length of the IU has a strong impact on phase and amplitudes of the waves measured using DAS. In practice, wavelengths shorter than one gauge length cannot be measured reliably, making it important that sufficiently short gauge lengths are selected for the target wavelength under consideration. While discussed in greater detail in the following paragraph, the authors note that very short gauge lengths, while possible to specify with some DAS IU's, will in general result in lower SNR \citep{dean_effect_2017}.

In summary, we recommend the following approach for selecting a DAS experiment's trace separation and gauge length. First, based on either prior experience with the IU or an on-site experimental test, the shortest gauge length capable of providing data of sufficient quality (i.e., high enough SNR) should be determined. Second, the shortest target wavelength of interest for the data acquisition should be determined based on the project's requirements. For example, for surface wave applications, the shortest target wavelength should be no more than approximately two-times the thickness of the near-surface layer one desires to resolve \citep{foti_guidelines_2018}. If the shortest desired wavelength is less than the minimum acceptable gauge length determined based on SNR, a compromise must be reached in terms of data quality and wavelength resolution. If the shortest desired wavelength is greater than the minimal gauge length, the experiment can proceed directly. Third, the channel separation should be selected to provide at least two channels per shortest wavelength to satisfy the Nyquist sampling theorem in space. Importantly, using small channel separations will not mitigate the issues of using a gauge length that is too long relative to the minimum target wavelength.

\section{MASW Processing Details}

We now discuss the procedure used for processing the surface wave dispersion data from the three arrays. To ensure a fair comparison between the geophone-derived and DAS-derived dispersion data, only the first 94 m of the 200-m long DAS arrays were used during MASW processing. As mentioned previously, three vibroseis chirps and five sledgehammer impacts were stacked, respectively, in the time-domain to produce a single vibroseis or sledgehammer waveform at each sensing location with higher SNR. Dispersion processing utilized both the FDBF with cylindrical-steering vector and square-root weighting and the frequency-wavenumber transform (FK) \citep{gabriels_situ_1987, nolet_array_1976}. We utilize two different wavefield transforms as a means to better account for epistemic uncertainty in surface wave dispersion processing and to allow for the calculation of more robust dispersion statistics \citep{vantassel_swprocess_2022}. To facilitate the extraction of multi-mode surface wave dispersion data for statistical comparison, a non-traditional, algorithmic approach was implemented for extracting surface wave phase velocity peak power points from the derived dispersion images. The algorithmic peak selection process involved finding all relative maximums in the dispersion image at each frequency. This is in opposition to the more commonly used approaches of manual peak power selection, or the selection of only the single phase velocity maximum at each processing frequency. This automated multi-peak procedure allows for a more complete utilization of the information present in the surface wave dispersion image, especially when multiple modes are present in the dispersion data, as is the case at the Hornsby Bend site. However, this procedure can also tend to produce greater sensitivity to the background noise in the dispersion images. As such, there is a trade-off when selecting the search's hyper-parameters to ensure that all meaningful peaks are extracted without the undue selection of spurious peaks caused by noise. The same peak selection process with the same hyper-parameters was used for all source, offset, and wavefield transformation combinations for both the geophone and DAS arrays.

\section{Extraction of High-Resolution, Multi-Mode Dispersion Data}

\begin{figure}[t]
	\includegraphics[width=\textwidth]{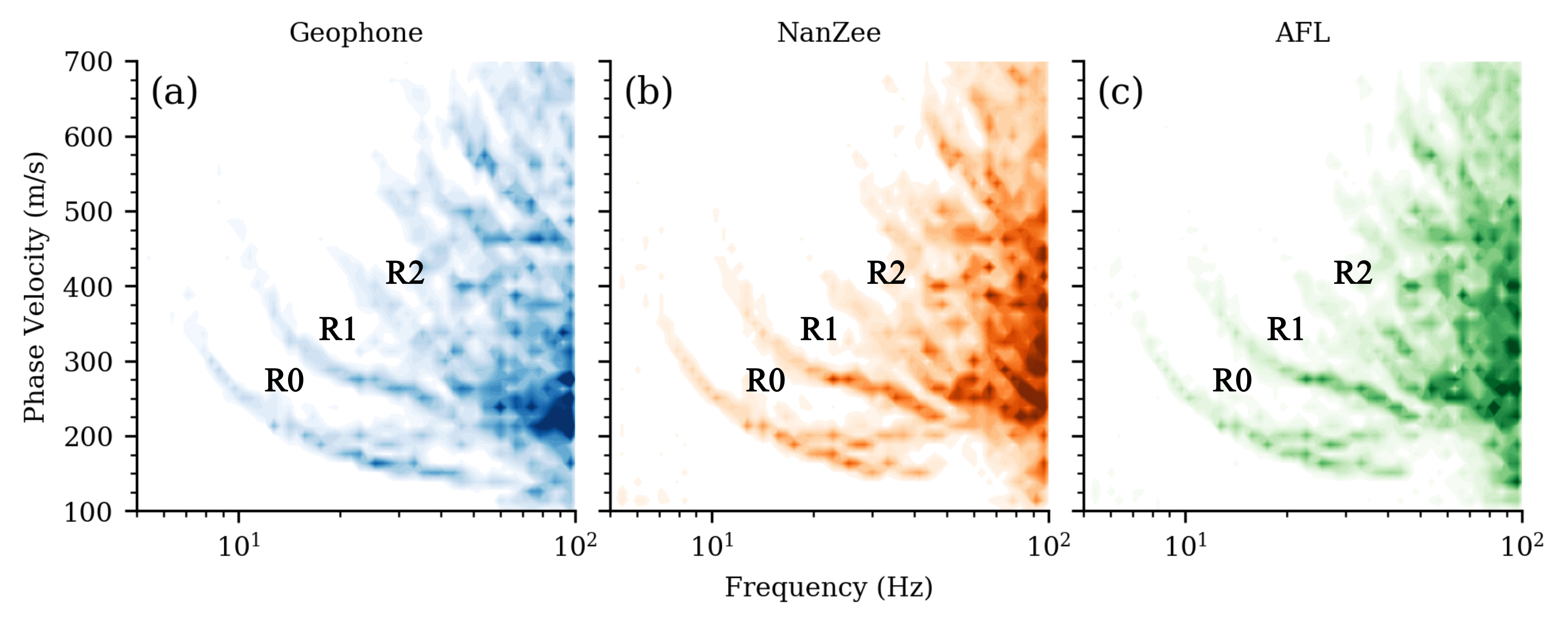}
	\caption{Visualization of the experimental surface wave dispersion data extracted from all source types, source positions, and wavefield transformations for the: (a) geophone array, (b) NanZee cable, and (c) AFL cable. The figure shows the raw experimental dispersion data binned into frequency-phase velocity pixels and plotted to better illustrate the density of the surface dispersion data. The darker regions of each image, which denote high experimental dispersion data density, denote regions of the frequency-velocity space close to the site's apparent Rayleigh wave modes. The first three Rayleigh wave modes of surface wave propagation are denoted in the figure with R0, R1, and R2, respectively. }
	\label{fig:8}
\end{figure}

The focus of this section is on the aggregation of surface wave dispersion estimates from the 32 source, offset, and wavefield transformation combinations, such that mean dispersion trends with accompanying statistical bounds can be compared between geophone and DAS data. To visualize the trends in the raw experimental dispersion data, the peaks in frequency-velocity space were binned into frequency-velocity pixels and plotted as a dispersion density plot in Figure \ref{fig:8}. Figure \ref{fig:8}a shows the aggregated dispersion results from the geophone array, Figure \ref{fig:8}b the NanZee cable, and Figure \ref{fig:8}c the AFL cable. The shading of the dispersion density plots denote the number of points in each pixel, with darker shading representing more frequency-velocity peaks per pixel. In all panels of Figure \ref{fig:8}, we can clearly observe the first three Rayleigh modes denoted as R0, R1, and R2, respectively. Furthermore, we observe excellent consistency between the three measurement systems. The R0 mode in all panels is clear and unambiguous. R1 and R2 are less clear than R0, particularly at frequencies above 50 Hz, however, R1 and R2 still represent clear mode trends.

Of particular note in Figure \ref{fig:8} is the wavelength limit of the coherent DAS-derived dispersion data. No dispersion data for R0 is available from either DAS cable below a wavelength of approximately 3 m, whereas the geophone R0 data extends up to approximately 2 m.  This observation, supports the previous discussion that there is a clear relationship between the minimum wavelength of surface wave dispersion data extracted from the wavefields and a DAS system's gauge length. In particular, we observe no coherent surface wave energy for any mode at wavelengths less than approximately two meters in the DAS wavefields, whereas we are able to clearly resolve such wavelengths using the geophone array. While the gauge length limitation of the DAS measurements is not particularly troublesome for this study, where a short gauge length of 2 m is used, this observation is critically important for those using DAS acquisition systems with larger gauge lengths (e.g., $\gg$ 2 m). Therefore, when using DAS systems for surface wave acquisition for engineering applications, where short wavelengths are critical for correctly resolving the stiffness of near-surface layers, the gauge length must be selected to be sufficiently short to permit good near-surface resolution. While the authors note that the high-frequency geophone-derived dispersion data are below the theoretical wavelength resolution limit based on the receiver spacing and spatial aliasing (i.e., $<$ 4 m), this limit can be relaxed when the measured dispersion data is of good quality and can be clearly differentiated from aliased dispersion data, as we do here \citep{foti_guidelines_2018}.

\begin{figure}[t]
	\includegraphics[width=\textwidth]{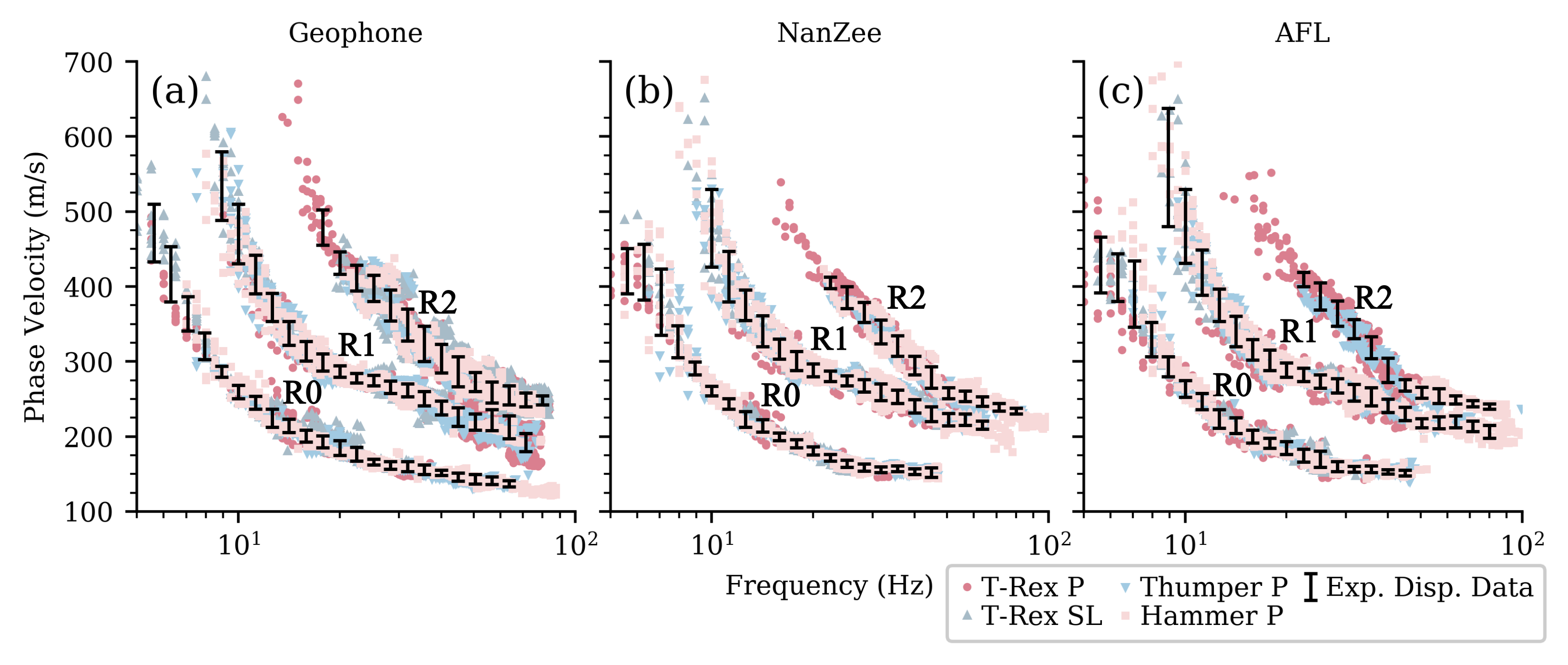}
	\caption{Statistical representation of the experimental dispersion data after performing interactive trimming to isolate the first three Rayleigh modes (i.e., R0, R1, and R2) for the: (a) geophone array, (b) NanZee cable, and (c) AFL cable. Each mode's statistical representation is denoted by error bars that represent the $\pm$ one standard deviation range. Those frequencies with fewer than five observations were neglected in the statistical calculations and therefore error bars are not shown.}
	\label{fig:9}
\end{figure}

The frequency-phase velocity peaks after performing interactive trimming are shown for the geophone array, NanZee cable, and AFL cable in Figure \ref{fig:9}a, \ref{fig:9}b, and \ref{fig:9}c, respectively. The interactive trimming process \citep{vantassel_swprocess_2022} involves the isolation of each continuous surface wave mode segment through a manual and somewhat subjective selection process. The reliability of the interactive-trimming process is dependent on the quality of the data and on the care and expertise of the analyst, and requires close consultation with the individual experimental dispersion images, such as those shown in Figure \ref{fig:5}, and the aggregated experimental dispersion data, such as that shown in Figure \ref{fig:8}, as well as a good understanding of the fundamentals of surface wave dispersion. Importantly, the purpose of interactive trimming is not to produce dispersion trends that are as clean as possible, but rather only to isolate the clear dispersive trends from spurious noise such that meaningful dispersion statistics can be calculated. Following the isolation of each surface wave mode using interactive trimming, dispersion statistics were calculated using a modified version of the workflow detailed by Vantassel and Cox (\citeyear{vantassel_swprocess_2022}) for surface wave dispersion processing. The modified dispersion statistics workflow first involves isolating each mode trend using interactive trimming. Once each mode has been isolated, duplicate phase velocity estimates from each dispersion observation (i.e., source-type, source-position, and wavefield-transformation combination) for the specific mode in question are discarded in lieu of the one with the highest relative wavefield power. The presence of duplicate observations of the same mode from a single dispersion image are a consequence of searching for all relative phase velocity peaks rather than solely the single phase velocity peak with the maximum power. The removal of these duplicate peaks ensures that each observation is treated equally and not unfairly weighted due to the presence of multiple phase velocity peaks in close proximity to one another. After removing duplicates from each observation, the continuous mode segments are interpolated to consistent frequencies on a logarithmic scale. These resampled observations (32 in total) are stored as rows in the data matrix. Each column of the data matrix is then used to calculate a mean and standard deviation at each of the resampled frequencies. Those frequencies with fewer than five observations were neglected in the statistical calculations. Note that while the authors believe that resampling in log-wavelength space generally produces superior results \citep{vantassel_swinvert_2021}, resampling in log-frequency was necessary to facilitate the multi-mode uncertainty-consistent inversions discussed later. The mean $\pm$ one standard deviation range of the resampled data in terms of log-frequency is what is shown with the dark error bars in Figure \ref{fig:9}a, \ref{fig:9}b, and \ref{fig:9}c. Note that the aforementioned data matrix was also used to calculate the correlation between the dispersion data measured at each frequency. These frequency-dependent correlations are part of the dispersion statistics that are required to implement the multi-mode, uncertainty-consistent surface wave inversion procedure discussed below. As correlation coefficient calculations require each variable (frequency in this case) to have the same number of observations, missing dispersion observations were replaced using mean imputation.

\begin{figure}[t]
	\centering
	\includegraphics[width=0.5\textwidth]{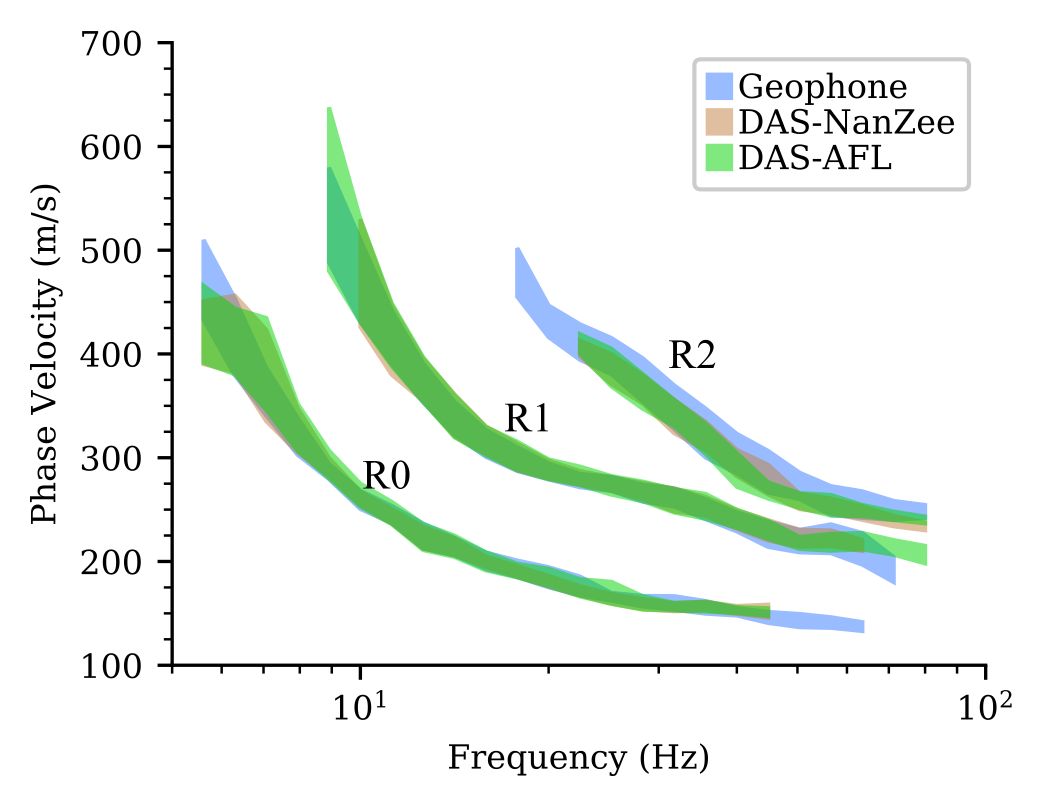}
	\caption{Comparison between the geophone-derived and DAS-derived (NanZee and AFL) experimental dispersion data at the Hornsby Bend site. The vertical range at each frequency represent the mean $\pm$ one standard deviation of the experimental dispersion data for the fundamental, first-higher, and second-higher Rayleigh modes (R0, R1, and R2, respectively).}
	\label{fig:10}
	\centering
\end{figure}

To facilitate a more direct comparison between the dispersion data extracted from each array, Figure \ref{fig:10} shows the mean $\pm$ one standard deviation range of the geophone-derived and DAS-derived experimental dispersion data from Figure \ref{fig:9} plotted directly on top of one another. Excellent agreement is observed between the three arrays, with the average difference in the mean and standard deviation across all three modes being less 5 \%. We note that at high and low frequencies the consistency between the geophone-derived and DAS-derived dispersion data decays slightly for all mode trends. This is due in part to the less clear dispersion trends at high and low frequencies that make consistent interactive trimming difficult (recall Figure \ref{fig:8}). We also note that the R2 mode derived from the geophones has a slightly higher mean trend and larger uncertainty than those estimated from the DAS cables. This is a direct result of a less clear R2 mode as observed using the geophone array (recall Figure \ref{fig:8}a) when compared to that derived from the DAS cables (recall Figure \ref{fig:8}b and \ref{fig:8}c). Regardless, the geophone-derived and DAS-derived experimental dispersion data are in excellent agreement, demonstrating that when appropriate considerations are made (i.e., proper cable selection, good cable-soil coupling, and sufficiently short gauge length) DAS can be used to measure surface wave dispersion data that is of equal quality to that acquired using geophones.

\section{Uncertainty-Consistent, Multi-Mode Surface Wave Inversion}

To further validate the quality of the DAS-derived dispersion data, we now seek to compare shear wave velocity (Vs) profiles obtained from surface wave inversions with ground truth information regarding the subsurface layering and stiffness from CPTU measurements made at the Hornsby Bend site (recall Figure \ref{fig:1}). We choose to only invert the DAS experimental dispersion data from the NanZee cable, based on the remarkable agreement between the experimental dispersion data from all three arrays (recall Figure \ref{fig:10}). In fact, the experimental dispersion data from all three arrays is in such excellent agreement that preliminary investigations showed that it was not possible to attain meaningfully different Vs profiles when properly accounting for inversion uncertainty. Importantly, the use of the NanZee experimental dispersion data should not be considered to be any endorsement or acknowledgement of the NanZee cable's superiority in any way, as the authors believe all three arrays provide practically identical experimental dispersion data. To perform the surface wave inversion, we extend the uncertainty-consistent procedure developed by Vantassel and Cox \citeyear{vantassel_procedure_2021} to multi-mode dispersion data. The reader will note that the uncertainty-consistent procedure by Vantassel and Cox \citeyear{vantassel_procedure_2021} was originally developed using only R0 data, however, the procedure's generality allows it to be easily extended to multi-mode dispersion data (this paper) and joint inversions (e.g., simultaneous inversion of Rayleigh and Love dispersion) (not shown here). However, the inversion of multiple modes (uncertainty-consistent or otherwise) is more challenging than inverting a single mode on its own because of the potential for modal inconsistencies between the 1D theoretical models and the observed experimental dispersion data, which may be the result of laterally variable materials underlying the array (i.e., non-1D conditions). To illustrate the consequence of modal inconsistency, consider the possibility that a dispersion image measured at a site contains a R0 and R1 trend and that the R0 trend is biased slightly due to experimental error such that it resides below the true R0 trend, whereas the R1 trend is less affected than R0 and resides close to the true R1 trend. Any attempt to simultaneously fit the R0 and R1 observations (or realizations of these observations in the uncertainty-consistent case) will result in modal inconsistencies that cannot be accounted for. Therefore, the resulting fit to the observations (or realizations) will be a best fit, in the least-squares sense, but is not guaranteed to be optimal in its fit to either R0 or R1. Naturally, the fitting process is further complicated when attempting to fit three modes simultaneously, as we do in this study. Nevertheless, using the data from the NanZee cable we were able to perform a multi-mode, uncertainty-consistent inversion to propagate the uncertainty quantified in the experimental dispersion data through the surface wave inversion process and into a resulting suite of 1D Vs profiles that represent, on average, the subsurface layering across the lateral extent of the array.

The multi-mode uncertainty-consistent inversion utilized four layering by number (LN) \citep{vantassel_swinvert_2021} parameterizations that contained 3, 5, 7, and 9 layers (i.e., LN3, LN5, LN7, and LN9, respectively) to account for inversion-derived uncertainty stemming from the inverse problem's non-uniqueness \citep{vantassel_procedure_2021}. Each LN parameterization was inverted using 250 unique realizations of the multi-mode experimental dispersion data, requiring 1,000 realizations in total (4 parameterizations * 250 realizations per parameterization). Each realization was inverted rigorously following the recommendations of Vantassel and Cox \citeyear{vantassel_swinvert_2021}. In particular, 100,000 trial models per trial inversion (i.e., 10,000 random, 300 iterations, and 300 models per iteration) and three trials per realization were performed. Therefore, each realization was searched with 300,000 trial models, thereby requiring 300 million models to be searched to develop the results presented here. The ability to perform surface wave inversion at this scale was provided via the high performance computing (HPC) application swbatch \citep{vantassel_jpvantasselswbatch_2021} that is publically available through the DesignSafe-CI \citep{rathje_designsafe_2017}. swbatch utilizes the global-search Neighbourhood Algorithm \citep{sambridge_geophysical_1999} as implemented in the Dinver module \citep{wathelet_surface-wave_2004} of the open-source software Geopsy \citep{wathelet_geopsy_2020} as its inversion engine. The computational time required to complete all 300 million dispersion forward calculations took fewer than 6 hours using 4 Skylake (SKX) nodes on the HPC cluster Stampede2 located at the Texas Advanced Computing Center (TACC).

\begin{figure}[t]
	\centering
	\includegraphics[width=0.5\textwidth]{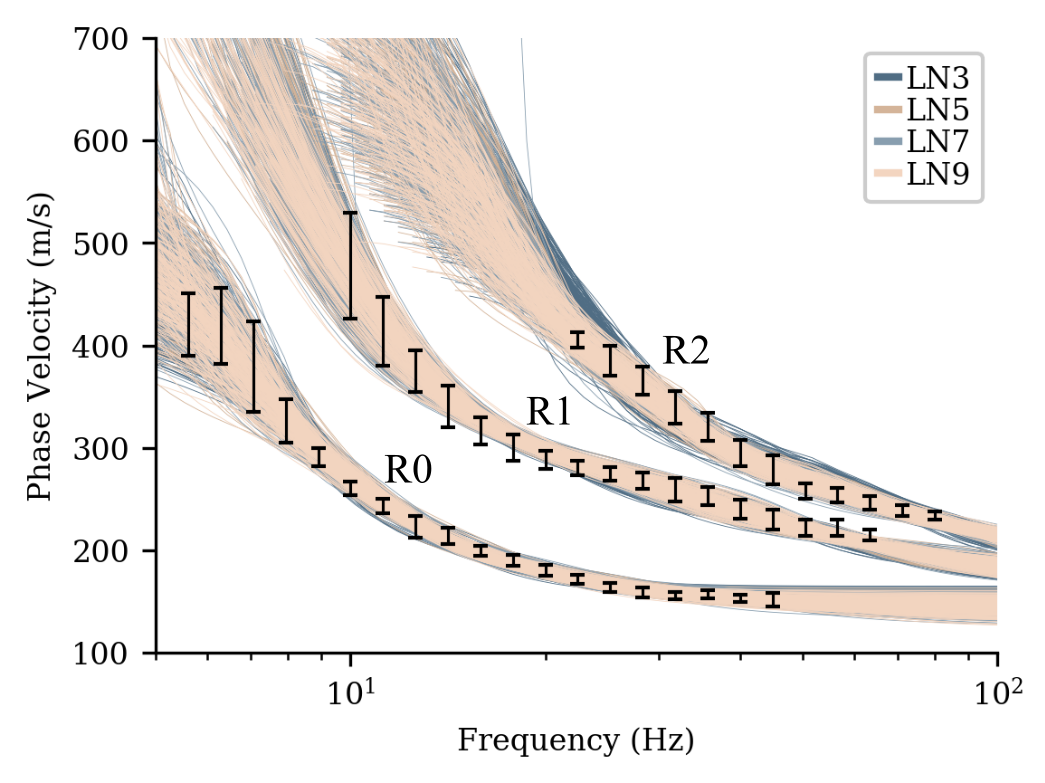}
	\caption{Comparison of the experimental dispersion data from the NanZee cable with the theoretical dispersion curves from the uncertainty-consistent, multi-mode inversion procedure. The theoretical dispersion curves are shown for each of the four layering by number (LN) parameterizations (i.e., LN3, LN5, LN7, and LN9). Each parameterization was used to invert 250 unique realizations of the fundamental, first-higher, and second-higher Rayleigh modes (R0, R1, and R2, respectively) to produce an ensemble of theoretical fits that are consistent with the uncertainty in the experimental dispersion data.}
	\label{fig:11}
	\centering
\end{figure}

The 1,000 theoretical dispersion curves recovered from the multi-mode uncertainty-consistent inversion process are shown alongside the experimental dispersion data from the NanZee cable in Figure \ref{fig:11}. In general, we observe an excellent match between the experimental and theoretical dispersion data over much of the dispersion bandwidth for all three Rayleigh modes. However, due to the modal inconsistencies discussed previously, it was not always possible to exactly match the experimental dispersion data's frequency-dependent uncertainty for all modes simultaneously. Examples of this are apparent in R1 around 60 Hz and again around 16 Hz, and in R2 around 60 Hz. Nonetheless, with the exception of these three locations, we observe that the inversion procedure was able to produce fits to the R0, R1, and R2 experimental dispersion data that appropriately accounts for each mode's frequency-dependent uncertainty. Note that the experimental dispersion data are shown with $\pm$ 1 standard deviation bounds, while the theoretical dispersion curves represent models that fit the full distribution.

\begin{figure}[t]
	\centering
	\includegraphics[width=0.5\textwidth]{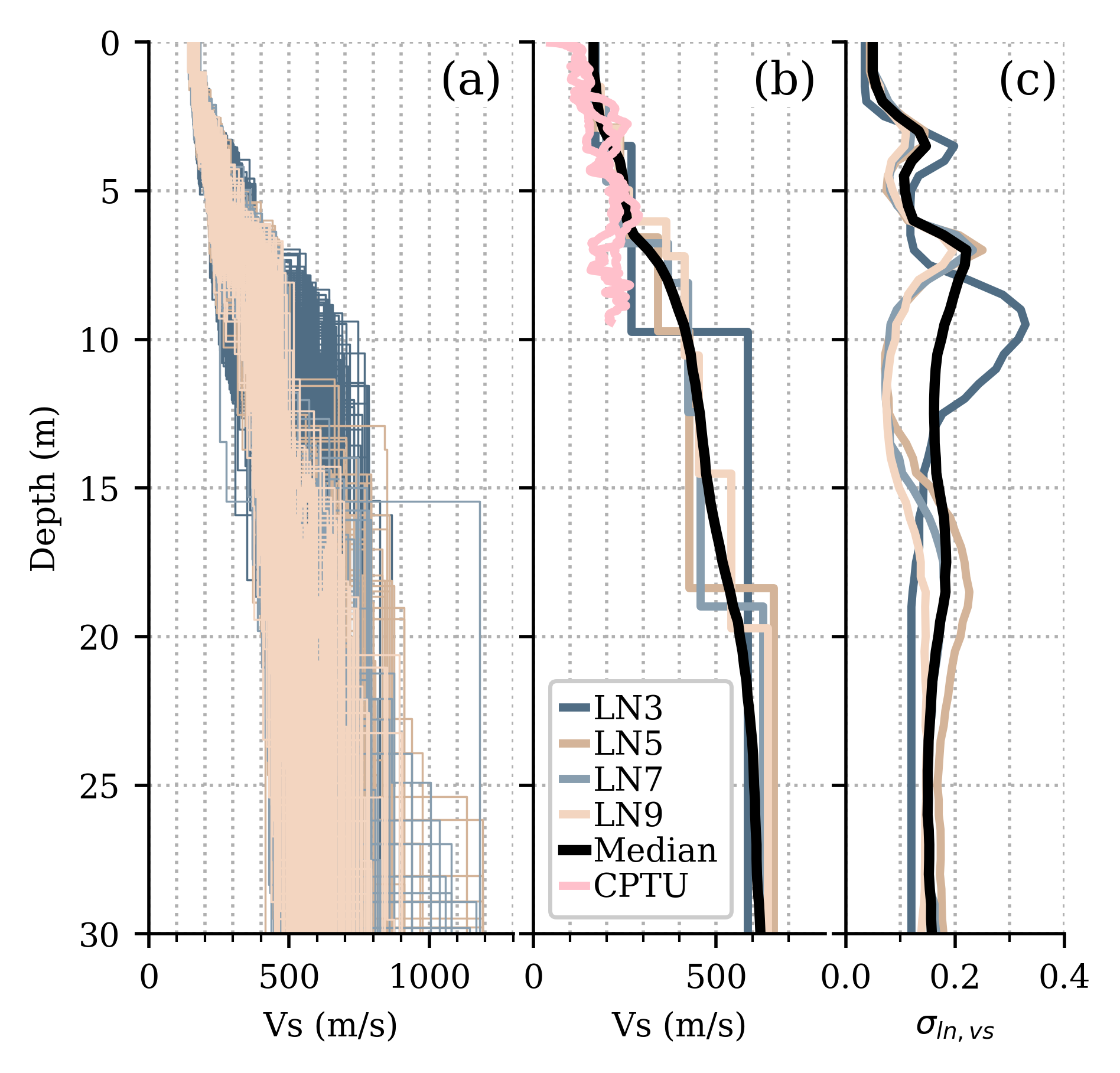}
	\caption{Shear wave velocity (Vs) profiles derived from the NanZee dispersion data using the uncertainty-consistent, multi-mode inversion procedure. Panel (a) shows the Vs profiles from the inversion of 250 realizations performed for each of the four LN parameterizations (i.e., 1,000 Vs profiles in total). Panel (b) shows the layer-by-layer median for each of the four LN parameterizations alongside the discretized median calculated across all LN parameterizations. Also shown in (b) are the CPTU measurements correlated to Vs. Panel (c) shows the intra- and inter-parameterization uncertainty as assessed using the lognormal standard deviation on Vs ($\sigma_{ln,Vs}$). Note the spikes in $\sigma_{ln,Vs}$ are due to the uncertainty in layer boundaries and a limitation of how $\sigma_{ln,Vs}$ has been calculated historically, and are not due to uncertainty in Vs directly.}
	\label{fig:12}
	\centering
\end{figure}

The 1,000 Vs profiles associated with the 1,000 theoretical dispersion curves from Figure \ref{fig:11} are shown in Figure \ref{fig:12}a. The uncertainty in the Vs profiles is the accumulation of the experimental uncertainty (i.e., the uncertainty in the experimental dispersion data) as well as the inversion-derived uncertainty (i.e., the uncertainty from the inverse problems itself) \citep{vantassel_procedure_2021}. To better illustrate the general trend of the four LN parameterizations, the layer-by-layer median profiles are shown in Figure \ref{fig:12}b alongside the discretised median calculated from all LN parameterizations. We chose to show the layer-by-layer medians for each parameterization (as opposed to a discretized median) to communicate the potential for sharp velocity contrasts when indicated by some of the inversion results (e.g., LN3 at 10 m depth). The discretized median is known to produce a smoothed approximation of the subsurface's velocity trend that may or may not be reasonable at all sites (Vantassel and Cox, 2021a). We believe that showing both the layer-by-layer median for each parameterization and the discretised median from all parameterizations illustrates two extreme, but plausible interpretations of the site's subsurface structure. Figure \ref{fig:12}b demonstrates excellent agreement between the layer-by-layer median Vs profiles and the discretized median profile for all LN parameterizations, which indicates a more gradual velocity increase with depth, except LN = 3, which indicates a more abrupt velocity contrast at about 10 m. We note here that each CPT was pushed to refusal, which also occurred at approximately 10 m depth. Figure \ref{fig:12}c shows the lognormal standard deviation of Vs ($\sigma_{ln,Vs}$) for each discretized LN parameterization (intra-parameterization uncertainty) and for all LN parameterizations (inter-parameterization uncertainty). With the exception of the near surface, where $\sigma_{ln,Vs}$  is quite small (~0.05), it is approximately 0.2 over the entire 30 m characterization depth. A value of $\sigma_{ln,Vs}$ of 0.2 is slightly higher than the 0.15 Stewart et al. \citeyear{stewart_guidelines_2014} recommended for sites with low variability, but substantially less than those previously recommended by others and routinely used in practice (e.g., 0.25 – 0.5) \citep{epri_seismic_2012, toro_probabilistic_1995}. Importantly, the spikes in $\sigma_{ln,Vs}$ are due to the uncertainty in layer boundaries and a limitation of how $\sigma_{ln,Vs}$) has been calculated historically and are not due to uncertainty in Vs directly.

To compare the inversion-derived Vs profiles to ground truth measurements made at the Hornsby Bend site, we present the CPTU measurements correlated to Vs using three different approaches from the literature. The CPTU-correlated Vs profiles are indicated in Figure \ref{fig:12}b. The three correlations are those developed by Hegazy and Mayne \citeyear{hegazy_global_2006}, Andrus et al. \citeyear{andrus_predicting_2007}, and Robertson \citeyear{robertson_interpretation_2009}. For the Andrus et al. \citeyear{andrus_predicting_2007} correlation we used the one proposed for Holocene-aged soils, as the surficial soil deposited at Hornsby Bend site have been dated by others to the early Holocene late Pleistocene \citep{blum_late_1994}. We note that, while we believe our decision to consider these soils as Holocene is a reasonable one, dating these soils as having been deposited in the Pleistocene would result in an increase to the estimated Vs by approximately 20 \% due to the inclusion of a soil-aging factor. We note that while Robertson \citeyear{andrus_predicting_2007} also states that in general Pleistocene soils have a Vs higher than Holocene aged soils with similar CPTU measurements, they opted to not include a soil-aging factor in their correlation because the reliable aging of soils on most projects is often infeasible. To present a single Vs profile for each CPTU, we averaged the Vs recovered by applying the three aforementioned correlations. In general we observe good agreement between the CPTU- and inversion-derived Vs values, with agreement being the most consistent over the top 6 m. Below 6 m, the CPTU-derived values agree more closely with the LN = 3 median Vs profile, which, as noted above, indicates a stronger impedance contrast at 10 m than the other inversion parameterizations containing more trial layers. This well illustrates one of the significant challenges associated with surface wave inversion; the experimental data can often be fit equally well using subsurface models with different numbers of layers, and unless the inversion layering can be constrained by a priori supporting information, one cannot know for sure which layering model is most representative of the true subsurface conditions \citep{cox_layering_2016}. Nevertheless, the near-surface Vs profiles derived using the uncertainty-consistent, multi-mode inversion procedure and the DAS dispersion data are in very good agreement with the ground-truth measurements made at the Hornsby Bend test site.

\section{Conclusions}

We compare DAS-derived and geophone-derived multi-mode dispersion data from active-source experiments using the MASW technique. In particular, we compare DAS data from two different tightly-buffered fiber-optic cables buried in a shallow trench with dispersion data extracted from horizontal geophones coupled to the ground surface using an aluminium spike. Wavefields with strong Rayleigh-type surface wave content were generated by using highly-controlled vibroseis sources and more-variable impact sources at four distinct source positions. We demonstrate that the use of frequency-dependent normalization of the dispersion image removes the effect of scaling, integration, and differentiation on the acquired waveforms, thereby mitigating the need to convert the measurements into consistent engineering units prior to comparing dispersion data. We show evidence that short wavelength DAS dispersion measurements are limited near and below the acquisition gauge length. These observations make gauge length selection an important factor to consider in future near-surface studies using DAS. To calculate experimental dispersion statistics representative for the Hornsby Bend site, all source type (4), source offset (4), and wavefield transformation (2) combinations (32 in total) were processed using a previously published MASW workflow. Multi-mode dispersion data was recovered from the experimental dispersion images by performing a non-traditional relative peak search, rather than the more common manual peak selection or absolute peak search techniques. While the use of a relative peak search allowed for the more complete extraction of information from the dispersion images, it also amplified the noise in the extracted dispersion data. Nonetheless, the fundamental, first-higher, and second-higher Rayleigh wave modes of propagation were able to be extracted using this procedure over a relatively broad frequency range for active-source studies (~6 to 90 Hz). Despite the gauge length limitation of the DAS system previously mentioned, the experimental dispersion data (mean $\pm$ one standard deviation range) recovered from the geophone and DAS systems show excellent agreement for all three recovered Rayleigh modes. The recovered multi-mode experimental dispersion data was inverted using a multi-mode, uncertainty-consistent procedure to recover suites of Vs profiles representative of the experimental and inversion-derived uncertainties. The uncertainty-consistent Vs profiles agreed favorably with CPTU tests available at the site. Therefore, when appropriate considerations are made to ensure proper cable selection, good cable-soil coupling, and sufficiently short gauge lengths, DAS can be an effective alternative to geophones for the purpose of acquiring dynamic signals for the intent of extracting high-resolution, multi-mode surface wave dispersion using the MASW technique.

\section{Acknowledgements}

This work was supported in part by the U.S. National Science Foundation (NSF) grants CMMI-2037900, CMMI-1520808, and CMMI-1931162. However, any opinions, findings, and conclusions or recommendations expressed in this material are those of the authors and do not necessarily reflect the views of NSF. Special thanks to Dr. Kevin Anderson at Austin Water – Center for Environmental Research for the access to the Hornsby Bend Biosolids Management Plant test site. Special thanks to Dr. Kenichi Soga for the contribution of the NanZee cable used in this study. Special thanks to Todd Bown and the OptaSense team for their assistance in configuring the ODH4, extracting and post-processing the seismic waveforms, and permitting us to publish these results. The raw seismic wavefield recordings made with the DAS and geophone arrays have been made publically available on the DesignSafe-CI \citep{vantassel_active-source_2022}. Active-source surface wave processing, interactive-trimming, and calculation of dispersion statistics was performed using the open-source, Python package \emph{swprocess} \citep{vantassel_jpvantasselswprocess_2021}. The multi-mode, uncertainty-consistent surface wave inversions were made possible by the HPC application, \emph{swbatch}, \citep{vantassel_jpvantasselswbatch_2021}. \emph{swbatch} is publically available on the DesignSafe-CI \citep{rathje_designsafe_2017}. Operations with the experimental dispersion data and the inversion results were performed using the open-source, Python package \emph{swprepost} \citep{vantassel_jpvantasselswprepost_2021}. The figures in this paper were created using Matplotlib 3.1.2 \citep{hunter_matplotlib_2007} and Inkscape 0.92.4.

\bibliographystyle{plainnat}
\bibliography{das_masw}

\end{document}